\documentclass[11pt, executivepaper]{article}
\usepackage[utf8]{inputenc}
\usepackage[T1]{fontenc}
\usepackage{natbib}
\usepackage{amsmath}
\usepackage{color}
\usepackage{amsfonts}
\usepackage{graphicx}
\usepackage[footnotesize, rm]{caption2}
\usepackage{enumitem}
\usepackage{geometry}
\usepackage{hyperref}
\hypersetup{colorlinks= true, allcolors=blue}
\setcitestyle{aysep={}}

\title{Is Quantum Mechanics Self-Interpreting?}

\author{Andrea Oldofredi\thanks{Contact Information: Universit\'e de Lausanne, Section de Philosophie, 1015 Lausanne, Switzerland. E-mail: Andrea.Oldofredi@unil.ch}}

\begin{document}

\maketitle 

\begin{abstract}
\cite{Fuchs2000} claimed that standard Quantum Mechanics needs no interpretation. In this essay, I show the flaws of the arguments presented in support to this thesis. Specifically, it will be claimed that the authors conflate QM with Quantum Bayesianism (QBism) - the most prominent subjective formulation of quantum theory; thus, they endorse a specific interpretation of the quantum formalism. Secondly, I will explain the main reasons for which QBism should not be considered a physical theory, being it concerned exclusively with agents' beliefs and silent about the physics of the quantum regime. Consequently, the solutions to the quantum puzzles provided by this approach cannot be satisfactory from a physical perspective. In the third place, I evaluate Fuchs and Peres arguments \emph{contra} the non-standard interpretations of QM, showing again the fragility of their claims. Finally, it will be stressed the importance of the interpretational work in the context of quantum theory.
\end{abstract}
\clearpage

\tableofcontents
\vspace{5mm}

\section{Introduction}

\emph{What is the correct interpretation of Quantum Mechanics?} This question not only is as old as QM itself, but also it still remains without a definite answer. Indeed, it is remarkable that nowadays there is no agreement about this issue among the experts working on the foundations of the theory, although a plethora of different proposals has been given. 

Looking at the history of quantum theory, in fact, it is possible to individuate several strategies to solve this conceptual riddle: many expanded the quantum formalism invoking hidden variables, or modified it introducing stochastic terms in the Schr\"odinger's equation referring to spontaneous random collapses of the wave function (see e.g. \cite{Durr:2013aa}, \cite{Bohm:1952aa}, and \cite{Ghirardi:1986aa}, \cite{Bassi:2003} respectively). Other believe that different results of quantum measurements are actualized in diverse, causally unrelated universes (\cite{Wallace:2012aa}), still others circumvent the problem changing the logic underneath quantum theory (\cite{French2010}, \cite{DallaChiara:2004}), while some authors propose a return to a neo-Copenhagenist view (\cite{Landsman:2017aa}, \cite{Ladyman2013}), \emph{etc}. Among this jungle of alternatives, nonetheless, there are eminent physicists affirming that standard quantum mechanics does not need an interpretation. \cite{Fuchs2000} indeed argued for the internal consistency of an ``interpretation without interpretation of QM'': nothing more than the bare formalism of quantum theory is needed, according to the authors, for ``using the theory and understanding its nature''. 

Although this paper is not extremely recent, it is still worth discussing it for three notable reasons: (i) it contains many subtleties concerning the issues of the interpretation of quantum mechanics - in general, it aims to explain the reasons for which people should not pay attention to the business of interpreting QM, in particular it provides arguments against non-standard interpretations of the theory -, (ii) it will give us the opportunity to analyze a rather peculiar position about the meaning of the quantum formalism, and (iii) its main moral still represents the position of the majority of physicists working with QM.
\vspace{2mm}

In the following sections I will discuss the arguments advanced by Fuchs and Peres to support their ``no-interpretation view''. I aim to show that their proposal is strongly interpreted, since the authors surreptitiously conflate QM with Quantum Bayesianism (QBism) - the most prominent subjective formulation of quantum mechanics. In the second place, I will explain the main reasons for which QBism should not be considered a physical theory, being concerned exclusively with agents' beliefs and silent about the ontology and physics of quantum objects dynamically evolving in spacetime. Rather, it must be taken as an epistemological framework able to weight the agents' subjective probabilities for measurement outcomes (as correctly pointed out in \cite{Jaeger:2009}, Chapter 3). Consequently, it will be argued that the solutions to the quantum puzzles advanced by the exponents of QBism can not be considered physically satisfactory. In the third place, I will evaluate Fuchs and Peres' arguments \emph{contra} non-standard interpretations of quantum theory, emphasizing the fragility of their claims. Finally, it will be stressed the relevance of the interpretational work in the context of the foundations of quantum physics.

\section{Quantum Mechanics \emph{without} Interpretation}

At the outset of their essay, Fuchs and Peres openly claim that the attention given by physicists and philosophers to the interpretation of standard quantum mechanics may lead people to a wrong impression concerning its validity and consistency; in fact, they reassure the readers right away affirming that ``[i]f quantum theory had been in crisis, experimenters would have informed us long ago!''. The main moral of their paper is unequivocal and can be summarized by saying that (i) standard quantum mechanics is consistent, being able to provide precise predictions for measurement outcomes - or more precisely, they claim it is worthless to worry about its interpretation to use it in actual experiments -, and (ii) its validity is well confirmed by empirical evidence. Thus, there is no urgency to investigate what is the correct interpretation of the quantum mechanical formalism: given the extraordinary empirical success of QM, its standard interpretation - or better, its bare formal structure - is sufficient to utilize the theory in practical situations and to understand ``its nature''. Consequently, the authors state, quantum theory does not need a particular interpretation, it is just self-interpreting. 

For the sake of clarity, it is worth stressing that the argumentative strategy employed by Fuchs and Peres concentrates exclusively on the empirical adequacy of quantum theory, without any attention to its physical content, i.e. the story QM tells about what are the objects and processes determining the observed measured outcomes to which they repeatedly allude.

In this precise regard, the authors state that to appreciate the physics of quantum theory, one has to reject every kind of philosophical demand for which QM should (somehow) provide a description of an external, mind-independent reality existing at the microphysical regime, or to give us an intimate knowledge of it. Consequently, one should not endorse non-standard interpretations of quantum mechanics, given that - in a way or another - these latter stress the importance of providing such a realistic description of the objects and processes taking place at the microscopic regime which are physically responsible for the observed measurements results. However, this desire of realism - inherited by a classical worldview which is difficult to overcome -, leads only to formally more complicated theories which do not improve the predictive power of quantum theory. The authors firmly claim that QM, on the contrary, does not describe reality, being an algorithm designed to calculate probabilities for macroscopic events, which are ``consequences of our experimental interventions''. Therefore, to search for a realistic description of the quantum world is a pointless effort denoting a dogmatic and conservative view about physics. Furthermore, there is no logical necessity, they conclude, that a physical theory should provide a realistic world-view of a specific domain of physical phenomena. 
\vspace{2mm}

In more detail, the authors argue that the wave function $\psi$, the central mathematical object of QM, represents only a formal expression of agents' degrees of belief, i.e. their available knowledge, about a certain experimental situation. Wave functions, therefore, do not refer to anything real in the world, ``in particular'', they wrote, ``no wave function exists before or after we conduct an experiment''. Consequently, QM is explicitly defined as a theory applicable exclusively to measurement situations, and not to physical phenomena independent from an experimental setting.
Moreover, this view about the nature of $\psi$ implies that the dynamical process described by the Schr\"odinger's Equation (SE) provides solely the evolution of the probabilities in the agent's mind. Hence, there is no motion of quantum objects in spacetime according to the authors. It is a straightforward logical consequence, then, that also the collapse of $\psi$ in measurement situations is not a physical process, i.e. it does not happen to quantum systems. QM, therefore, it is not anymore a mechanical theory in the usual physical sense - i.e. a theory which specifies its fundamental objects and how they dynamically behave and interact in spacetime -, but it becomes a branch of probability theory, concerning exclusively agents' degree of belief - or human knowledge - in experimental situations. 
\vspace{2mm}

To conclude the section, it is possible to summarize and characterize Fuchs and Peres theses about quantum theory, i.e. their interpretation without interpretation of quantum mechanics, as follows:
\begin{enumerate}
 \item QM does not describe an external physical reality at microscopic regimes;
 \item QM provides algorithms to compute probabilities of macroscopic events;
 \item The wave function exclusively represents agents' degrees of belief about experimental situations;
 \item The dynamical laws of the theory obeyed by the wave function represent the evolution of agents' probability assignments.
\end{enumerate}

\section{Quantum Mechanics or Something Else?}

Having expressed Fuchs and Peres positions, two questions arise: 
\begin{enumerate}
\item Is standard quantum mechanics really internally consistent? 
\item Do these theses faithfully represent the physical content of quantum mechanics?
\end{enumerate}
 
\noindent The former is carefully answered by \cite{Norsen:2008} who, in response to Fuchs and Peres, correctly point out that QM cannot be considered internally coherent for ($a$) it is plagued by the Measurement Problem (MP), and ($b$) because the dynamical laws governing the behavior of quantum systems, the Schr\"odinger equation and the collapse postulate, are mutually inconsistent. To this regard, it should be also stressed that the physical processes responsible for the collapse of the wave function - or equivalently, for the suppression of the Schr\"odinger evolution - are not described by standard quantum theory, with the consequence that the interactions occurring in measurements processes - which give raise to experimental outcomes - do not receive a physically meaningful explanation within the context of QM. Thus, the projection postulate, although efficient, seems to be a pragmatical rule introduced \emph{ad hoc} in order to tame inconsistencies with respect to observed experimental facts. In fact, as correctly pointed out by Maudlin, quantum mechanics does not provide any satisfactory clarification of what distinguishes a measurement interaction from a non-measurement interaction:
\begin{quote}
[w]hat the traditional theory did \emph{not} do is state, in clear physical terms, the conditions under which the non-linear evolution takes place. [...] But if the linear evolution which governs the development of the fundamental object in one's physical theory occasionally \emph{breaks down} or \emph{suspends itself} in favor of a radically different evolution, then it is a physical question of the first order exactly under what circumstances, and in what way, the breakdown occurs. The traditional theory papered over this defect by describing the collapses in terms of imprecise notions such as ``observation'' or ``measurement'' \citep[p. 9]{Maudlin:1995aa}.
\end{quote}
Finally, on the one hand it must be underlined that the stochastic collapse of the wave function introduces an essential, arbitrary and not precisely defined demarcation between the micro- and macroscopic regimes (see  \cite{Curiel:2009} Section 6 on this point), on the other, the notions of ``measurement'' and ``observer'', albeit pivotal within the axioms of standard QM and taken as unexplained primitive concepts, are neither mathematically nor physically well-defined, i.e. there are no variables in the equations of the theory referring to these notions (for a detailed analysis of these issues the reader may refer to the seminal book \cite{Bell:2004aa}). 

It goes without saying that Fuchs and Peres may readily reply something along the following lines: if $\psi$ does not represent anything real, then SE and the collapse postulate cannot be inconsistent, not describing physical processes taking place in space and time. More precisely, these authors would maintain that the collapse postulate has a clear meaning: after the performance of a particular experiment, the agent just sees, observes what outcome has been produced, and consequently updates her knowledge concerning the measurement situation at hand - which evolved continuously in time according to the Schr\"odinger's dynamics prior the observation. It is specifically this process of updating the observer's knowledge which causes the suppression of the Schr\"odinger's evolution. The crucial consequence of this hypothetical response is that the MP is simply dissolved together with the other objections mentioned a few lines above and logically related to it. This fact brings us to the second question, namely, whether what Fuchs and Peres claim about QM is actually what the theory says. Simply put, the correct answer is a plain ``No''.

To motivate such conclusion, it must be primarily stressed that to identify the standard formulation of quantum mechanics with the above mentioned theses means to surreptitiously conflate two highly different interpretations of the theory. What  Fuchs and Peres actually presented in their paper is Quantum Bayesianism (QBism), a subjective interpretation of quantum theory mainly developed by Caves, Fuchs and Schack (see \cite{Fuchs2002}, \cite{Fuchs2007}, \cite{Fuchs2010}, \cite{Fuchs2014} and references therein), which must be clearly distinguished from the standard version of the theory (see notably \cite{Sakurai1994} for a detailed presentation). Usually supporters of QBism rely on (i) anti-realist readings of Bohr and the young Heisenberg, and (ii) on the subjective interpretation of the quantum state endorsed by Pauli and the late Heisenberg, in order to claim that the main theses of the QBist's view are indeed shared by the fathers of the Copenhagen interpretation of QM. However, the subjective interpretation of the wave function advocated by Pauli and the late Heisenberg \emph{is not} considered part of the \emph{standard} contemporary presentations of quantum theory, where $\psi$ is statistically interpreted - more precisely, where $|\psi|^2$ is interpreted as the probability density to find a particle in a given volume if a measurement of position were performed. Remarkably, it is also debatable whether or not these subjective positions can be considered part of the Copenhagen interpretation of QM, given the internal heterogeneity of positions held by its supporters (to this regard see \cite{Howard:2004}, where it is questioned even if we can properly speak of a unitary Copenhagen interpretation). Secondly, the instrumentalist and positivist attitude of Fuchs and Peres (and more generally the attitude of QBists) is not fully endorsed by Bohr (see  \cite{Jaeger:2009}, pp. 124-136 for textual evidence) and Heisenberg, who explicitly claimed that 
\begin{quote}
``[t]he positivists have a simple solution: the world must be divided into that which we can say clearly and the rest, which we had better pass over in silence. But can any one conceive of a more pointless philosophy, seeing that what we can say clearly amounts to next to nothing. If we omitted all that is unclear, we would probably be left with completely uninteresting and trivial tautologies'' (\cite{Heisenberg:1971} p. 213).
\end{quote}

\noindent Thus, it seems unwarranted and unjustified to identify the theses maintained by Fuchs and Peres with the principles of quantum mechanics. Contrary to QBism, according to standard quantum theory wave functions do represent physical objects (as elementary particles, atoms and molecules) which dynamically evolve in space and time obeying the two processes of QM: the SE when a system is not measured, and the collapse postulate when a measurement occurs. Therefore, the theory explicitly aims to provide a formal description of the physics taking place at the quantum length scales. Consequently, standard QM is absolutely \emph{not} about agents' degrees of belief, and its dynamical laws do not refer to the evolution in time of subjective judgements on experimental outcomes. 

Pointed out this fact, we are now in the position to answer the questions posed at the outset of this section:

\begin{enumerate}
\item Standard quantum mechanics is not consistent for the above mentioned arguments;
\item Fuchs and Peres were neither presenting, nor speaking about standard quantum theory, they confused it on purpose with QBism.
\end{enumerate}
\vspace{2mm}

Then, let us now evaluate this Bayesian interpretation of the quantum formalism.\footnote{For the sake of precision, it should be underlined that QBism is not a solipsistic theory: it needs the existence of a macroscopic reality external to the agents' minds for its foundations - the theory is instead antirealist concerning the reality of microphysical objects. Moreover, QBism is not ruled out by the PBR theorem since it does not presupposes that the wave function represents a physically ``real'' state of a quantum system, as clarified by \cite{Ben:2017}, p. 80. For other critical studies on QBism see notably \cite{Jaeger:2009}, Chapter 3, \cite{Marchildon:2015} and \cite{Timpson:2008}.} The main reasons to support this perspective, according to its advocates, are the following: it is a local theory, it is not ontologically committed to the existence of peculiar quantum objects, dissolves the measurement problem and the inconsistency between SE and the collapse postulate. 

Although these are certainly positive features for a quantum theory, some qualifications seem necessary in this case. In the first place, QBism is trivially local since it does not describe physical processes in spacetime, but only the evolution of agents' beliefs about experimental outcomes. Thus, non-local correlations do \emph{not} represent in this context a peculiar behaviour of quantum objects in space. As a consequence, the local character of QBism is irrelevant to reconcile quantum theory with the physics of special relativity, given that QBism does not provide a physical account of the objects and processes taking place at the quantum length scales. Being concerned exclusively with agents' knowledge, QBism is naturally defined as an efficient epistemological - or better operational - theory able to weight agents' bets, or beliefs about measurement results. Nonetheless, the foundational, physical issue about the compatibility between special relativity and QM is left untouched by this framework. Thus, the local character of quantum Bayesianism has no relevant \emph{physical} meaning. Secondly, for the very same reasons QBism \emph{does not solve} the ontological conundrums of QM, for it is simply not concerned with the behavior of physical objects moving and interacting in spacetime.\footnote{It is worth stressing that QBism cannot be straightforwardly considered an instrumental reading of quantum mechanics. In fact, many exponents of quantum Bayesianism not only claim that classical physics does describe a macroscopic reality, i.e. it provides a realistic description of our classical regime, but also, they state that it is the aim of science to truly describe the world, an aim to which QM contributes only indirectly as pointed out by \cite{Healey:2016}. Moreover, they claim that statements of QBism - contrary to a genuine form of instrumentalism - do have a peculiar truth-functional content since ``quantum state assignment is true or false relative to the epistemic state of the agent assigning it, insofar as it corresponds to that agent's partial beliefs concerning his or her future experiences (beliefs the agent should have adopted in accordance with the Born Rule). But what makes this quantum state assignment true or false is not the physical world independent of the agent'' (\cite{Healey:2016}). Thus, on the one hand, it is not possible to quickly dismiss QBism as an instrumental interpretation of quantum theory, on the other, QBists dissolve the conundrums of the quantum mechanical description of physical phenomena claiming that QM simply does not provide a realistic description of the world at the quantum length scales, endorsing a sort of middle ground position between realism and instrumentalism.}
In the third place, one has to underline that QBism - if considered a physical theory - would lack in explanatory power. With an ontology of agents' beliefs QBism would show an inherent lack of capacity to explain the interference phenomenon obtained, for instance, in the double-slit experiment. It would not be able to explain the predictions for the quantization of energy from the Schr\"odinger's equation, since the latter refers to the evolution in time of agents' beliefs. It would neither provide a justification  for the existence of anti-matter as a consequence of the Dirac's equation. Examples of this sort can be generated \emph{ad infinitum}. In addition, if we would consider QBism a physical theory, it would be not clear how from its ontology we could meaningfully describe the physical processes causally responsible for observed measurement outcomes. Nor we would be able to explain how the classical, macroscopic reality would emerge from a microscopic, quantum scale - in other words, it is not clear how a theory of the classical limit could be obtained. 

Finally, although the supporters of QBism often speak about the ``information'' possessed by an agent about a specific measurement situation, their notion of information is radically different from that employed by the quantum information approach as \citep[p. 233]{Jaeger:2009}, p. 233 underlines:
\begin{quote}
[t]he Informational interpretation differs fundamentally from the Radical Bayesian interpretation, which suggests that the fundamental referent of quantum information theory is [human] knowledge, in that the Informational interpretation considers quantum mechanics to describe \emph{objectified} information, with a metaphysics more closely resembling logical atomism, and does not deny quantum mechanics the status of a physical theory. 
\end{quote}

\noindent This remark is essential since it must be clearly stated that the success of quantum information science should not be considered evidence in favor of QBism.
\vspace{2mm}

These considerations bring us to a simple conclusion: one should not consider QBism a physically satisfactory interpretation of quantum mechanics. It should be at best characterized, instead, as an empirically successful epistemological or probabilistic theory. Therefore, it is deceptive and misleading (i) to conflate QBism with standard QM - or, even worse, to sell the quantum Bayesian theses as claims made or shared by the supporters of the Copenhagen interpretation -, and (ii) to maintain that such a framework can clarify the puzzles of quantum theory, since it does not provide any improvement in our understanding of the \emph{physical} solutions to its foundational issues.

\section{Evaluating the Arguments Against Non-Standard Interpretation of QM}

In this section we will evaluate the three main arguments given by Fuchs and Peres against the non-standard interpretations of QM:

\begin{enumerate}
  \item There is no logical necessity for QM to provide a realistic worldview of the physics at quantum length scales;
  \item Non-standard interpretations of quantum theory do not improve its predictive power;
  \item Non-standard interpretations introduce extra-structures to obtain a classical description of the world, but QM does not describe physical reality.
\end{enumerate}

Concerning (1), it is a plain historical fact that realistic quantum theories are obtainable in several ways, thus, there is equally no logical necessity for them to be un-obtainable. Therefore, (1) is too weak in order to dismiss the importance of non-standard interpretations of QM. The second argument is simply not correct since the Ghirardi-Rimini-Weber (GRW) theory\footnote{This argument applies also to the formulations of GRW theory implementing a mass density and a flash ontology.} provides predictions which are slightly different with respect to those of standard quantum mechanics. Consequently, GRW can be tested experimentally against it. Therefore, (2) cannot be generalized to every non-standard interpretation of quantum theory. On the other hand, QBism does not offer any new prediction whatsoever; thus, Fuchs and Peres' proposal is victim of their own argument. Furthermore, although Bohmian Mechanics is empirically equivalent to QM, it provides new tools in several practical applications, especially in the fields of quantum engineering or quantum chemistry as explicitly showed in \cite{Oriols:2012aa}. Thus, improving the applicability of quantum theory in several cases, it can be claimed that BM is a compelling framework also from a practical perspective - which is not the case of QBism.

The third argument misses the point of realistic quantum theories. These frameworks introduce additional structures to overcome the technical and conceptual difficulties affecting the formalism QM, and to improve its explanatory power. For instance, the additional structures of BM and of GRWm, GRWf are meant to solve the quantum measurement problem, to provide a consistent dynamics to the theory and to eliminate physically ill-defined notions - e.g. measurement and observer - from the axioms of quantum mechanics (see notably \cite{Bell:2004aa}, \cite{Bassi:2003}, \cite{Durr:2013aa}). Moreover, these theories are dynamically highly non-classical, and in many cases their ontology is completely different with respect to classical physical theories.

In conclusion, these three arguments neither support the claim that QM does not need an interpretation, nor question the robustness of the currently available non-standard interpretations of quantum theory. 

\section{Closing Remarks: The Importance of the Interpretational Work}

In this essay it has been firstly argued that Fuchs and Peres surreptitiously conflated QBism with QM in order to support their ``no-interpretation'' view. Secondly, I showed not only that this is a methodological incorrect move, but also emphasized the evident limitations of QBism, if considered a physical theory. Finally, I explained the flaws in Fuchs and Peres' arguments against the non-standard interpretations of QM.
\vspace{2mm}

In conclusion, let me stress the reasons for which interpretational work is still essential within the foundations of quantum physics. In the first place, it has to be stressed the standard formulation of QM is neither internally consistent, given the incoherence between its dynamical laws, nor it is ontologically well-defined, since its axioms notoriously contain ill-defined notions. Furthermore, QM is affected by the measurement problem. These technical and conceptual problems, if not treated, will be \emph{by construction} inherited in more fundamental physical theories structurally based on the quantum formalism as Quantum Field Theory and quantum gravity, as masterly pointed out in \cite{Barrett:2014aa}, who shows that the measurement problem is equally present in relativistic quantum theory. 

In the opinion of the present author, viable solutions to the quantum puzzles are offered by realist quantum theories such BM or GRW, since these frameworks provide meaningful descriptions and explanations of the objects and processes which are causally - and physically - responsible for measurement outcomes. These theories achieve such results without any inconsistency among their dynamical laws, and without any ill-defined notion appearing in their axioms. Furthermore, being equipped with a primitive ontology\footnote{Considering a physical theory $T$, its primitive ontology corresponds to the fundamental objects postulated by $T$. The expression ``primitive ontology'' has been introduced in \cite{Durr:2013aa}, Chapt. 2. For a careful presentation of this notion, the reader may consider \cite{Allori:2008aa}, \cite{Allori:2013aa}, \cite{Esfeld:2014ac} and references therein.}, they are able to yield rigorous explanation for the emergence of the operator algebra characterizing the formal structure of QM. More precisely, \cite{Durr:2004c} and \cite{Goldstein:2012} showed how the observables of quantum theory are reduced to the primitive ontologies and dynamics of BM and GRW theories respectively. Finally, it is also worth noting that GRW theories admit relativistic extensions, while several interesting models for QFT have been proposed in the context of Bohmian Mechanics. Therefore, these frameworks - albeit still well-developed only at the non-relativistic regime - should be considered consistent possible alternative strategies to overcome the problems inherent to the standard interpretation of quantum theory.

In sum, in order to have ontologically and formally well defined quantum theories at different length/energy scales, one must carefully settle the interpretational issues of standard QM and try to resolve its inconsistencies, contrary to what Fuchs and Peres claimed. 
\clearpage

\bibliographystyle{apalike}
\bibliography{PhDthesis}
\end{document}